\documentclass[conference, a4paper,UTF8]{IEEEtran}
\IEEEoverridecommandlockouts
\usepackage{fancyhdr}
\usepackage{color}
\usepackage{amssymb}
\usepackage{amsmath}
\usepackage{amsmath,amssymb,amsfonts}
\usepackage{algorithmic}
\usepackage{graphicx}
\usepackage{textcomp}
\usepackage{amsmath,amssymb,amsfonts}
\usepackage{algorithmic}
\usepackage{graphicx}
\usepackage{float}
\usepackage{subfigure}
\usepackage{setspace}
\usepackage{hyperref}
\usepackage{textcomp}
\usepackage{amsfonts}
\usepackage{paralist}
\usepackage{tikz}
\usetikzlibrary{shapes.arrows}
\usepackage{amsmath}
\usepackage{algorithm}
\usepackage{algorithmic}
\usepackage{amsthm}
\usepackage{warpcol}
\usepackage{booktabs}
\usepackage{multirow}
\usepackage{graphicx}
\usepackage{booktabs}
\usepackage{rotating}
\usepackage{placeins}
\usepackage{mathrsfs}
\usepackage{ifsym}
\usepackage{tikz}
\usepackage{subfigure}
\usepackage{geometry}
\usetikzlibrary{decorations.pathreplacing}
\usetikzlibrary{decorations.pathreplacing,calligraphy}
\usepackage{makecell}

\usepackage{lastpage}
\usepackage{setspace}
\usepackage{algorithm} %format of the algorithm
\usepackage{algorithmic} %format of the algorithm
\usepackage{multirow} %multirow for format of table

\usepackage{amsmath,bm}
\usepackage{xcolor}
\usepackage{layout}
\usepackage{amsmath}
\usepackage{hyperref}
\usepackage{color}
\usepackage{amsfonts}
\usepackage{graphicx}
\begin{document}
\begin{spacing}{1.0}

\title{On the Submodularity of Diffusion Models: Equivalent Conditions and Applications}
\author{
\IEEEauthorblockN{Fangqi Li, Wenwen Xia, Chong Di}
\IEEEauthorblockA{\textit{School of Cyber Science and Engineering,} \\
\textit{Shanghai Jiao Tong University,}\\
Shanghai, China. \\
\IEEEauthorblockN{\url{solour_lfq@sjtu.edu.cn}}}
}
\date{Today}
\maketitle
\begin{abstract}
The diffusion model has been a crucial component in studies about social networks. Many studies, especially these about influence maximization concern the proof of the submodularity of particular diffusion models. Such proofs have been model-dependent and are somewhat ad hoc. In this paper, we prove a theorem that provides a necessary and sufficient condition for a diffusion model to be submodular. This theorem can be used to justify the submodularity of an arbitrary diffusion model. We also apply this theorem to build a projection operator that maps an arbitrary diffusion model into a submodular one. Moreover, we use the established theorem to propose a diffusion model of multiple heterogeneous pieces of information that partially features submodularity.


\end{abstract}
\section{Introduction}
\label{section:1}
\emph{Diffusion model} is an important part of studies concerning social networks and the propagation pattern of information within, it formalizes and answers this question: in what way does a piece of information propagate in a complex community? A community is a collection of participants that interacts with each other, and its properties are usually embedded in a weighted directed graph $\mathcal{G}=\left(\mathcal{V},\mathcal{E},\mathcal{W} \right)$, given $\mathcal{G}$, a diffusion model computes what subset of participants can be influenced by some information released by another subset of participants. For example, in the problem of influence maximization, we are to select a subset $\mathcal{S}$ of $\mathcal{V}$ with no more than$K$ participants, such that an information propagation begins from it can affect as many participants as possible. The influenced subset in this scenario is denoted by $\sigma(\mathcal{S})$ and we are to maximized $|\sigma(\mathcal{S})|$. Here the propagation function $\sigma(\cdot)$ is given by a specific diffusion model.

Problems like influence maximization are innately combinatorial optimizations, whose space of possible solutions has the size $\mathcal{O}(|\mathcal{V}|^{K})$ which grows exponentially with $K$, so efficient algorithms are hardly feasible. To reduce the cost of calculation in such problems, [1] proposed that: if we take advantage of the submodularity of diffusion model and the theorems in [2], then the greedy algorithm (during which $K$ nodes that maximize the marginal propagation increment are selected sequentially) can result in a propagation range which is no smaller than $\left(1-\frac{1}{e}\right)$ of the theoretical optimal propagation range. Studies hitherto [3][4][5] utilize other graph features or extra data structures to boost the calculation and have derived diversified and fruitful results. However, the proof of the submodularity of an arbitrary diffusion model remains challenging and inspiring, [1] proved the submodularity of two specific diffusion models, namely \emph{independent cascade model} (IC) and \emph{linear threshold model} (LT), but what about a general diffusion model? Moreover, does there exist a simple and feasible equivalent condition for a diffusion model to be submodular? For non-submodular diffusion models, is it possible to approximate them using submodular ones? What properties does this approximation method hold? At last, for information propagation in complex scenarios as competitive propagation or cooperative propagation, can we design diffusion models that partially feature submodularity? We try to provide answers to these questions in this paper. The contributions of this paper are three-folded:
\begin{enumerate}
\item We show that under some trivial assumptions, there exists a necessary and sufficient condition for a diffusion model to be submodular.

\item We propose a projection operator that transforms an arbitrary diffusion model into a submodular diffusion model. This operator is a legal mapping under a certain definition of equivalence between diffusion models.

\item The established results are utilized to coin a diffusion model of multiple pieces of information within a network. This model partially features submodularity and can be applied in further studies concerning competitive influence maximization.

\end{enumerate}

This paper proceeds as follows: Section \ref{section:2} reviews the formulation of diffusion model, together with other elementary knowledge. Section \ref{section:3} contains our main theorems about the equivalent condition of the submodularity of diffusion models. Section \ref{section:4} extends the dissusion in Section \ref{section:3} and provides some practical corollaries. Section \ref{section:5} concludes this paper.

\section{Preliminaries}
\label{section:2}
\subsection{Formulation of the Network}
The diffusion model formulates the information propagation within a network, the network here is a rather abstract and broad concept that can be instantiated into a social network, computer network, connected water body [4], etc. Therefore related researches have a broad range of possible applications. The network is usually formulated by a weighted directed graph $\mathcal{G}=\left(\mathcal{V},\mathcal{E},\mathcal{W} \right)$, where the weight of an edge $w_{e},e=(u,v)\in \mathcal{E}\subset \mathcal{V}^{2}$ reflects the intensity of the influence that node $u$ exerts on node $v$ (in this paper we use the terminologies \emph{node}, \emph{participant} and \emph{vertex} interchangeably.) The physical interpretation of such intensity is model-dependent. It could be a function of probability, a relative measure of weight corresponding to some threshold metric, etc. In practice, it is impossible to provide a $\mathcal{G}$ without specifying a corresponding diffusion model which explains the meaning of its weights.

\subsection{Formulation of the Diffusion Model}
Given the network structure $\mathcal{G}$, a diffusion model $M$ defines a conditional probability measure on the power set of vertices:
$$\mathcal{P}_{M}(\mathcal{T}|\mathcal{S}),\mathcal{S}\subset\mathcal{T}\subset\mathcal{V},$$
whose physical interpretation is: the probability that information propagation begins from $\mathcal{S}\subset \mathcal{V}$ terminates in $\mathcal{T}\subset \mathcal{V}$. To derive the theorems below, we exert two assumptions on diffusion models:
\begin{enumerate}

\item \textbf{Markovian}: We assume that for any vertex $v\in \mathcal{V}$, whether or not it is influenced/activated (in this paper, being \emph{influenced} or \emph{activated} means the same thing) is solely determined by the state of its parent nodes $S_{v}$, and is independent from the activation state of other vertices or the order of activation (by the \emph{state} of its parent nodes $S_{v}$ we mean the binary encoding of $S_{v}$, where $0$ means a vertex is unactivated/does not receive the influence and $1$ denotes otherwise). To put it in other words, the event of activation of $v$ can be embedded by a conditional probability $\mathcal{P}_{M}(v'|S'_{v})$, where $S'_{v}$ is a specific activation state of $S_{v}$ with altogether $2^{|S_{v}|}$ possible values (we also exert an order on all possible states of $S_{v}$, to do so, we firstly define an arbitrary order on $S_{v}$ and encode every state $S'_{v}$ using binary code so every $S'_{v}$ can be mapped to a unique number between $0$ and $(2^{|S_{v}|}-1)$, hence a natural order is feasible), while $v'$ is a binary variable that reflects whether vertex $v$ is activated or not.

\item \textbf{Acylic}: We assume that in propagation, the graph $\mathcal{G}$ can be taken as an acyclic graph, hence there exists a topological order $\prec$ on $\mathcal{V}$. If we sort $\mathcal{V}$ using $\prec$, then for any vertex $v$, all its parent nodes $u\in S_{v}$ appear before $v$ [6]. In other words, the unfolding process of the propagation can be conducted according to this topological order, for all the necessary information that determined whether vertex $v$ can be activated or not are ready when $v$ is visited according to $\prec$.
\end{enumerate}

Under these two assumptions, the propagation induced by a diffusion model in $\mathcal{G}$ from $\mathcal{S}\subset \mathcal{V}$ to $\mathcal{T}\subset\mathcal{V}$ can be seen as proceeding as follows:
\begin{enumerate}
\item Exerting a topological order $\prec$ on $\mathcal{V}$.
\item Marking the vertices in $\mathcal{S}$ as activated ones.
\item Iterating over $\mathcal{V}$ according to $\prec$, for each $v\in\mathcal{V}$, determing whether or not it would be activated by sampling from $\mathcal{P}_{M}(v'|S'_{v})$.
\item Finally, the probability $\mathcal{P}_{M}(\mathcal{T}|\mathcal{S})$ is given by the product of all conditional probabilities on vertices:
$$
\begin{aligned}
\mathcal{P}_{M}(\mathcal{T}|\mathcal{S})=\prod_{v\in\mathcal{V}}&\mathcal{P}_{M}(v'=1|S_{v'})^{\mathbb{I}[v\in\mathcal{T}]}\\
\cdot&\left(1-\mathcal{P}_{M}(v'=1|S_{v'})\right)^{1-\mathbb{I}[v\in\mathcal{T}]}.
\end{aligned}
$$
where $\mathbb{I}[...]$ is an boolean indicator function.
\end{enumerate}
The product in the equation above is taken according to $\prec$ so each term is ready to be computed according to the two assumptions given before.

\subsection{Submodularity}
A function that define on the power set of $\mathcal{V}$, $f:2^{\mathcal{V}}\rightarrow \mathbb{R}$ is said to be \textbf{submodular} if $\forall \mathcal{S}\subset \mathcal{T}\subset\mathcal{V}, \forall v\in\mathcal{V}/\mathcal{T}$ the following inequality holds:
\begin{equation}
\label{equation:1}
f(\mathcal{S}\cup \left\{v \right\})-f(\mathcal{S})\geq f(\mathcal{T}\cup \left\{v \right\})-f(\mathcal{T}).
\end{equation}
Submodularity is an elegant and desirable property that might be enjoyed by the propagation range of a diffusion model. Once the submodularity is established, the influence maximization problem with respect to the corresponding diffusion model can be readily solved by greedy methods and a result lower bounded by $\left(1-\frac{1}{e} \right)$ of the theoretical optimum is secured. Therefore the proof of the submodularity of a diffusion model (i.e., the submodularity of the range of propagation according to this diffusion model) is of vital significance.

Most diffusion models include randomness in the propagation process (so the conditional probability $\mathcal{P}_{M}(\mathcal{T}|\mathcal{S})$ hardly degenerates), therefore the propagation range as a scalar function of the set of seeds $\mathcal{S}$ can be formulated as:
\begin{equation}
\label{equation:2}
f(\mathcal{S})=\mathbb{E}\left[|\sigma_{M}(\mathcal{S})|\right]=\sum_{\mathcal{T}\subset\mathcal{V}}\mathcal{P}_{M}(\mathcal{T}|\mathcal{S})\cdot |\mathcal{T}|.
\end{equation}

\section{Main Theorems}
\label{section:3}
\subsection{Transforming to a New Measure}
The submodularity of a function as $\eqref{equation:2}$ is difficult to analyze, so we resort to a transformation in measure, formally, consider the following form:
\begin{equation}
\label{equation:3}
f(\mathcal{S})=\sum_{G}\mathcal{P}_{M}(G)\cdot f(G,\mathcal{S}).
\end{equation}
In which we try to transform the randomness in $P_{M}(\mathcal{T}|\mathcal{S})$ into another auxiliary variable $G$, and hope that $G$ is independent from $\mathcal{S}$ but is capable of representing the distribution of propagation introduced by $M$. According to our discussion in Section 2.2., the propagation process is a cascade of application of activation rule for each $v$ and its $S_{v}$ \emph{per se}, by activation rule we mean a reinterpretation of $\mathcal{P}_{M}(v'|S'_{v})$, whose every instance is tantamount to a rule of activation that assigns each activation pattern/state of $S_{v}$ (recall that there are altogether $2^{|S_{v}|}$ possibilities) a deterministic result about the state of $v$. So we can denote a specific response pattern of $v$ by $G_{v}$, a binary vector of length $2^{|S_{v}|}$, whose the $i$-th component denotes whether $v$ would be activated or not under the state of $S_{v}$ encoded binarily as $(i-1)$. The collection of all possible $G_{v}$ corrseponding to all $v$ forms $G$. Section 2.2. has justified that $\mathcal{P}(G)$ preserves all information in $\mathcal{P}_{M}(\mathcal{T}|\mathcal{S})$, so we can study \eqref{equation:3} instead of \eqref{equation:2} without loss of generality [7]. Particularly, the following relationships hold, which might help to delve into this transformation between measures and variables:
$$\mathcal{P}_{M}(G_{v})=\prod_{i=1}^{2^{|S_{v}|}}\mathcal{P}_{M}\left(v'=G_{v}(i)|S'_{v}\sim (i-1)\right),$$
$$\mathcal{P}_{M}(G)=\prod_{v\in\mathcal{V}}\mathcal{P}_{M}(G_{v}),$$
where $G_{v}(i)$ denotes the reaction of $v$ according to $G_{v}$ under the activation state $S'_{v}$ and is a binary variable, $0$ for unactivation and $1$ for activation, $S'_{v}\sim(i-1)$ denotes that $S'_{v}$, an activation state of $S_{v}$, is binarily encoded as $(i-1)$. Since the $\mathcal{P}_{M}(v'|S'_{v})$ for each $v$ are independent from each other, there is no particular numeric constraints on $\mathcal{P}_{M}(G)$ except for normalization. Finally, the deterministic propagation function $f(G,\mathcal{S})$ conditioned on $G$ can be evaluated as follows: iterating over the topological sequence on $\mathcal{V}$ and judging whether $v\in\mathcal{V}$ is activated or not according to $G_{v}\in G$, finally the number of activated vertices is returned. Note that since $G_{v}$ has collected all possible states of $S_{v}$, so $G$ is independent from $\mathcal{S}$. It is now obvious that given the set of seeds $\mathcal{S}$, \eqref{equation:2} and \eqref{equation:3} introduce the same distribution of activated nodes among $\mathcal{V}$.

\subsection{The First Main Theorem}
If for any possible $G$, $f(G,\cdot)$ appears to be a submodular function, then $f(\cdot)$, as a convex combination of submodular functions, turns out to be a submodular function. To check for this point, pluggin \eqref{equation:1} into \eqref{equation:3} and it is straightforward to see:
$$
\begin{aligned}
&\ \ \ f(\mathcal{S}\cup\left\{v \right\})-f(\mathcal{S})\\
&=\sum_{G}\mathcal{P}_{M}(G)\left[f(G,\mathcal{S}\cup\left\{v \right\})-f(G,\mathcal{S}) \right]\\
&\geq \sum_{G}\mathcal{P}_{M}(G)\left[f(G,\mathcal{T}\cup\left\{v \right\})-f(G,\mathcal{T}) \right]\\
&=f(\mathcal{T}\cup\left\{v \right\})-f(\mathcal{T}).
\end{aligned}
$$
Unfortunately, there exists some assignment of $G$ whose corresponding combination of activation rules refutes submodularity. Thus there is only a strict subset of all possible $G$ on which submodularity holds. Therefore, if it turns out that a diffusion model $M$ can be transform in a way such that all probability on this specific subset sums up to one, then $f(G,\cdot)$ would always be submodular with respect to this diffusion model and \eqref{equation:3} would be submodular. Hence it is crucial to identify the class of $G$ on which submodularity holds. Formally, we prove the following theorem:

\emph{Theorem 1. For a diffusion model $M$, $\forall v\in\mathcal{V}$, let $v$'s expectations of being activated under altogether $2^{|S_{v}|}$ states of activation of $S_{v}$ be collected in a vector $\textbf{a}_{M}(v)$ of size $2^{|S_{v}|}$, where $S'_{v}$ are sorted using binary code order. If $\forall v\in\mathcal{V}$, the following equation holds:
\begin{equation}
\label{equation:4}
\exists \textbf{b},\textbf{a}_{M}(v)=\textbf{M}\textbf{b},
\end{equation}
where $\textbf{b}$ is a probability vector of length $2^{|S_{v}|}$, whose components are non-negative and sum up to unity. And the second-rank tensor $\textbf{M}$ is defined by:
\begin{equation}
\label{equation:5}
\textbf{M}_{i,j}=\bigvee_{m=1}^{|S_{v}|}\left(i(m)\land j(m)\right),
\end{equation}
where $i(m),j(m)$ denotes the $m$-th index of the binary code for $(i-1),(j-1)$ respectively. $\textbf{M}$ depends on $|S_{v}|$ while $\textbf{b}$ depends on $v$. Then $M$ is submodular in a sense that each $G$ it supports introduces a submodular propagation range. The reverse theorem also holds.}

\emph{Proof:} We first prove that this condition is sufficient for submodularity. Note that \eqref{equation:5} collects all $G_{v}$ of the following category: firstly, $v$ is connected to $S_{v}^{*}$, a subset of $S_{v}$, secondly, if at least one vertex in $S^{*}_{v}$ is activated then so is $v$, otherwise $v$ is not going to be activated. The pattern of connection is encoded in the dimensionality of $i$, while the activation state of $S_{v}$ is encoded in the dimensionality of $j$. Only if there exists at least one activated vertex in $S_{v}$ and is connected to $v$ will $v$ be activated. Here \emph{there exists at least one} and \emph{activated and is connected to} are formally reflected in \eqref{equation:5} by $\bigvee$ and $\land$. If the expectation to be activated at $v$ under different states (i.e. $\textbf{a}_{M}(v)$) satisfies \eqref{equation:4}, then it would be safe to transform the behavior of $M$ at $v$ by sampling from these $G_{v}$ (the connection-activation pattern) by probability $\textbf{b}$. Since all the components of $\textbf{b}$ sum up to one, the behavior of $M$ is restricted to this subset of $G$. Under this condition, the proof in [1] can be revoked to show submodularity. A piece of information that begins from $\mathcal{S}$ affects $v$ if and only if there is an activated path that connects one vertex in $\mathcal{S}$ to $v$, the activation of each path is controlled by $\textbf{b}$. In this way, $M$ would turn out to be submodular.

Reversely, if the condition $\textbf{a}_{M}(v)=\textbf{M}\textbf{b}$ fails to be met, then $\textbf{a}_{M}(v)$ can be spanned on another set of basis $\textbf{M}'$, which is different from $\textbf{M}$ in \eqref{equation:5}, $\textbf{a}_{M}(v)=\textbf{M}'\textbf{b}'$, then the propagation is tantamount to activating $v$ according to a rule $G_{v}$ different from \eqref{equation:5}'s semantics. Explicitly, the activation rule at $v$ can no longer be \emph{always} written as a disjunctive clause of a subset of $S_{v}$. Therefore at least a conjunctive clause is included, i.e., $v$ would be activated only if all vertices in a subset $S^{**}_{v}$ of $S_{v}$ (with strictly more than one node) are activated simultaneously. Otherwise $v$ can not be activated (since $\textbf{M}\textbf{b}$ defined under \eqref{equation:4} and \eqref{equation:5} has exhausted the complimentary events). Now any conjunctive rule appears to be contradictive against submodularity (intuitively, in \eqref{equation:1}, conjunctive rules enable a larger $\mathcal{T}$ to cooperate with $v$ so potentially more activation conditions are to be met, hence submodularity breaks down.) So once $M$ has to be transformed so it might have to be spanned on this non-submodular basis, the submodularity of $M$ is no longer preserved. Practically, it is no longer safe to sample from all space of $G$ supported by $M$ since a subset of it might introduce non-submodularity propagation. This completes the proof of the theorem.

To be illustrative, consider a case where $|S_{v}|=3$, so there are altogether 8 different activation states: \{000,001,010,011,100,101,110,111\}. Under this symbolization, the matrix $\textbf{M}$ takes the form:
$$\textbf{M}=
\begin{pmatrix}
0 & 0 & 0 & 0 & 0 & 0 & 0 & 0 \\
0 & 1 & 0 & 1 & 0 & 1 & 0 & 1 \\
0 & 0 & 1 & 1 & 0 & 0 & 1 & 1 \\
0 & 1 & 1 & 1 & 0 & 1 & 1 & 1 \\
0 & 0 & 0 & 0 & 1 & 1 & 1 & 1 \\
0 & 1 & 0 & 1 & 1 & 1 & 1 & 1 \\
0 & 0 & 1 & 1 & 1 & 1 & 1 & 1 \\
0 & 1 & 1 & 1 & 1 & 1 & 1 & 1 \\
\end{pmatrix}.
$$
To have a sight into the semantics of $\textbf{M}$, consider for example $\textbf{M}_{3,5}$, the binary code of the row index and the column index is \{011\} and \{101\} respectively, using the same construction in the proof of the theorem, they are tantamount to the activation state of $S_{v}$ as:
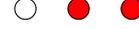
\begin{figure}[htbp]
\centering
\begin{tikzpicture}[scale = 0.7]
\draw [black, fill=white] (0,0) circle [radius = 0.2];
\draw [black, fill=red] (1,0) circle [radius = 0.2];
\draw [black, fill=red] (2,0) circle [radius = 0.2];
\end{tikzpicture}
\caption{The third activation state of $S_{v}$, red marks activation, white marks unactivation.}
\label{figure:1}
\end{figure}
\\While the connection looks like:
\begin{figure}[htbp]
\centering
\begin{tikzpicture}[scale = 0.7]
\draw (0,0)--(1,-1) [black, thick] ;
\draw (1,0)--(1,-1) [black, thick][dashed] ;
\draw (2,0)--(1,-1) [black, thick] ;
\end{tikzpicture}
\caption{The fifth connection state of $S_{v}$, dashed line marks unactivation.}
\label{figure:2}
\end{figure}
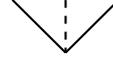
\\Combine figure.\ref{figure:1} and figure.\ref{figure:2} yields:
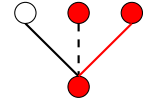
\begin{figure}[htbp]
\centering
\begin{tikzpicture}[scale = 0.7]
\draw [black, fill=white] (0,0.2) circle [radius = 0.2];
\draw [black, fill=red] (1,0.2) circle [radius = 0.2];
\draw [black, fill=red] (2,0.2) circle [radius = 0.2];
\draw [black, fill=red] (1,-1.2) circle [radius = 0.2];
\draw (0,0)--(1,-1) [black, thick] ;
\draw (1,0)--(1,-1) [black, thick][dashed] ;
\draw (2,0)--(1,-1) [red, thick] ;
\end{tikzpicture}
\caption{The evaluation of $\mathbf{M}_{3,5}$.}
\label{figure:3}
\end{figure}
\\It can be observed that $\textbf{M}_{3,5}$ takes the value 1 since there is an activated vertex that happens to be connected to $v$. By assuming $\textbf{a}_{M}(v)=\textbf{M}\textbf{b}$, the diffusion model' s effect at $v$ can be spanned onto only conjunctive rules as illustrated.

The proof for IC model and LT model in [1] can be included as special cases of Theorem 1, since IC model is corresponding to a specific assignment of $\textbf{b}$:
$$b_{S^{*}_{v}}=\prod_{i=1}^{|S_{v}|}p_{i}^{\mathbb{I}[v_{i}\in S^{*}_{v}]}(1-p_{i})^{(1-\mathbb{I}[v_{i}\in S^{*}_{v}])},$$
where $p_{i}$ are parameters in IC model. It is easy to verify that $\textbf{b}$ is a probability vector, since it exhausts the space of pairwise independent connection. While LT model corresponds to the following $\textbf{b}$:
$$b_{v_{i}}=p_{v_{i}},b_{\emptyset}=1-\sum_{i=1}^{|S_{v}|}w_{v_{i}},$$
$$\text{else }b_{\cdots}=0,$$
where $w_{v_{i}}$ are parameters in LT model. It can be concluded that IC model and LT model cover two extreme cases in the class of all submodular diffusion models given by Theorem 1. IC model assumes the components of $\textbf{b}$ are constrainted by independency, while LT model only normalized on a small subset of $\textbf{b}$ (those corrsepond to only one activated connection edge). For general $\textbf{b}$, more degrees of freedom is left for the diffusion models that satisfy the two prerequsites in Section 2.2.

\subsection{The Second Main Theorem}
Theorem 1 takes the advantage of auxiliary variable $G$ to algebraically express $M$, we can further demonstrate that the submodularity of a diffusion model in propagation range is equivalent to the submodularity of any individual vertex in activation expectation. Formally, we prove the following theorem:

\emph{Theorem 2: A diffusion model $M$ is submodular if and only if for any $v\in\mathcal{V}$, the activation expectation $E_{M}^{v}$ as a function of type: $2^{S_{v}}\rightarrow [0,1]$ (it maps a subset of $S_{v}$ into the expectation/probability of $v$'s being activated) is non-negative, monotonic and submodular.}

\emph{Proof: } We only need to show that the condition in Theorem 2 is equivalent to that in Theorem 1. For $v\in\mathcal{V}$, given the activation state of $S_{v}$ be $S^{*}_{v}$, the expectation/probability of $v$ being activated is $E_{M}^{v}(S^{*}_{v})$ by definition. According to Theorem 1, this value is also the $\left(S^{*}_{v(2)}+1\right)$-th component (decoding this binary expression) of $\textbf{a}_{M}(v)$, i.e., the inner product of the $\left(S^{*}_{v(2)}+1\right)$-th row of $\textbf{M}$ and $\textbf{b}$ ($S_{v(2)}^{*}$ denotes the value of $S_{v(2)}^{*}$ as a binary code.)

The non-negativity is conjugately implied by both definitions.

For monotonity, if $S^{*}_{v}\subset S^{**}_{v}$ then \eqref{equation:5} declares that compared with the $\left(S^{*}_{v(2)}+1\right)$-th line of $\textbf{M}$, the $\left(S^{**}_{v(2)}+1\right)$-th line replaces some 0s with 1s, but not a single 1 is converted to 0. Therefore compared with $E_{M}^{v}(S^{*}_{v})$, more components of $\textbf{b}$ are added into $E_{M}^{v}(S^{**}_{v})$, therefore the monotonity holds.

For submodularity, consider the value of:
$$E_{M}^{v}(\mathcal{S}\cup\left\{u\right\})-E_{M}^{v}(\mathcal{S}),$$
the value of which is the summation of the probability of this kind of $G_{v}$: such a $G_{v}$ connects $u$ and $v$, meanwhile, none vertex in $\mathcal{S}\subset S_{v}$ is connected to v. For$\mathcal{S}\subset\mathcal{T}\subset\mathcal{S}_{v}$, the value of:
$$E(\mathcal{T}\cup\left\{u\right\})-E(\mathcal{T})$$ sums up the probability of that $G_{v}$ connects $u$ and $v$, while fails to connect any vertex in $\mathcal{T}\subset S_{v}$ to $v$. The second set of $G_{v}$ is obvious a subset of the first one, since $\mathcal{S}\subset\mathcal{T}$, so a $G_{v}$ that connects $u$ and $v$ while leaves $\mathcal{T}$ and $v$ disconnects necessarilly disconnects $\mathcal{S}$ from $v$. Therefore the summation of the probability of the second set is no larger than that of the first. Hence the proof the Theorem 2 is finished.

Theorem 2 is symmetric and elegant in a sense that it shows: the submodularity of the propagation range is tantamount to the submodularity of the activation expectation of any vertex. This is a non-trivial observation since these two kinds of submodularity is different in nature. The submodularity of expectation for any vertex implies that the diffusion model can be intactly spanned on a set of submodular auxiliary variables, which further guarantees the submodularity of the diffusion model.

\section{Applications}
\label{section:4}
In this section, we apply the two theorems in Section. \ref{section:3} to yield some interesting results. First, note that different from the works in [1] where different ways of proving are applying to different diffusion models, it is possible to uniformly apply Theorem 1 and Theorem 2 to any diffusion model given its Markovian and acyclic properties.

\subsection{Submodular Projection Operator}
Considering this problem: given a non-submodular diffusion model, can we convert it to a submodular one while preserving its propagation property as well? We can solve this problem by applying Theorem 1, note that for fixed $|S_{v}|$, the set:
\begin{equation}
\label{equation:n1}
C=\left\{ \textbf{M}\textbf{b}:\textbf{b}\text{ is a probability vector, \textbf{M}}\sim\eqref{equation:5}\right\}
\end{equation}
\\is convex, therefore if we minimize a convex error function as 2-norm, we ends up with a unique global optimum [8]:
\begin{equation}
\label{equation:6}
\arg\min_{\textbf{a}^{*}\in C}\left\{|\textbf{a}_{M}(v)-\textbf{a}^{*}|_{2}^{2} \right\}.
\end{equation}
Now for an arbitrary diffusion model, we can conduct the projection in \eqref{equation:6} to every $v\in\mathcal{V}$ (if at a particular $v$ \eqref{equation:4} holds, then the projection degenerates to identity) and we end up with a submodular diffusion model that is close to the original one. This projection operator is a legal mapping that yields a unique image since \eqref{equation:6} returns a unique result. Formally, an arbitrary diffusion model $M$ can be projected into a submodular one $M^{*}$ whose activation expectation at each $v$ is:
\begin{equation}
\label{equation:7}
\textbf{a}^{*}(v)=\arg\min_{\textbf{a}^{*}(v)\in C(v)}\left\{|\textbf{a}_{M}(v)-\textbf{a}^{*}(v)|_{2}^{2} \right\},
\end{equation}
where $\textbf{a}_{M}(v)$ is given by the original $M$, while $C(v)$ as \eqref{equation:n1} depends solely on $|S_{v}|$. Here we consider two diffusion models $M_{1}$ and $M_{2}$ are equivalent if and only if:
$$\forall v\in\mathcal{V},\textbf{a}_{M_{1}}(v)=\textbf{a}_{M_{2}}(v).$$

\subsection{Diffusion Model for Multiple Information}
When multiple pieces of information are transmitted simultaneously in the network. The whole scenario becomes much more complicated and hard to analyze. Despite some reported works [9][10][11], there is still no consensus on a diffusion model for multiple pieces of information.

The inherent problem within a diffusion model of multiple information is that the activation expectation vector/function takes value in a product space (we have to consider a vertex's response to different information at the same time.) So there is no direct generalization of order, not to mention submodularity.

So far the best we can do in carrying submodularity to diffusion models for multiple information is to consider submodularity for each kind of information. We propose a partial linear model in that different types of information enjoy submodularity in propagation range. Taking two types of information $I_{1}$ and $I_{2}$ for example. The idea behind is to adopt Theorem 2 and assume that for each $v\in \mathcal{V}$, its expectation to be activated by $I_{1}$ or $I_{2}$ is a submodular function of vertices activated by $I_{1}$ or $I_{2}$ in $S_{v}$. To ensure that such submodularity holds for both $I_{1}$, consider that:
\begin{equation}
\label{equation:8}
E_{1}^{v}(\mathcal{S}\cup\left\{u \right\})-E_{1}^{v}(\mathcal{S})\geq E_{1}^{v}(\mathcal{T}\cup\left\{u\right\})-E_{1}^{v}(\mathcal{T}),
\end{equation}
where $E_{1}^{v}$ is a function that computes the probability that $v$ being activated by $I_{1}$, $\mathcal{S}\subset\mathcal{T}$ and $u\in S_{v}/\mathcal{T}$. Now assume that for each $S^{*}_{v}\subset S_{v}$ activated by $I_{1}$, let $S_{v}/S^{*}_{v}$ be activated by $I_{2}$. Then it is intuitive to assume that
\begin{equation}
\label{equation:9}
E_{2}^{v}(S_{v}/S^{*}_{v})=1-E_{1}^{v}(S^{*}_{v}).
\end{equation}
Combine \eqref{equation:8} and \eqref{equation:9}:
\begin{equation}
\label{equation:10}
E_{2}^{v}(\mathcal{T}'\cup\left\{u \right\})-E_{2}^{v}(\mathcal{T}')\geq E_{2}^{v}(\mathcal{S}'\cup\left\{u\right\})-E_{2}^{v}(\mathcal{S}'),
\end{equation}
where $\mathcal{T}'=S_{v}/(\mathcal{S}\cup\left\{u \right\})$, $\mathcal{S}'=S_{v}/(\mathcal{T}\cup\left\{u \right\})$ so $\mathcal{S}'\subset \mathcal{T}$. However, by assumption the reverse of \eqref{equation:10} has to held by the submodularity of $I_{2}$, so it turns out that:
\begin{equation}
\label{equation:11}
E_{2}^{v}(\mathcal{T}'\cup\left\{u \right\})-E_{2}^{v}(\mathcal{T}')= E_{2}^{v}(\mathcal{S}'\cup\left\{u\right\})-E_{2}^{v}(\mathcal{S}').
\end{equation}
For $I_{1}$ and $E_{1}^{v}$, an analogous proposition holds. A naive realization of \eqref{equation:11} is to assume that the increment of each vertex in $S_{v}$ for either $E_{1,2}^{v}$ is independent of any other vertex. In this case the diffusion model is similar to LT model. To summarize, our \textbf{Parital Linear Model for Multiple Information} (PLMMI) operates as follows, assuming that there are $N$ different types of information:
\begin{enumerate}
\item For $S_{v}$ of $v\in\mathcal{V}$, each $u\in S_{v}$ is assigned $N$ weights $\left\{w_{u,n}\right\}_{n=1}^{N}$ such that $\sum_{u\in S_{v}}\sum_{n=1}^{N}w_{u,n}=1.$
\item Upon propagating, a vertex $v\in\mathcal{V}$ computes its expectation to be activated by any type of information $I_{n}$ by summarizing over the weights of vertices activated by $I_{n}$, i.e.,
$$E_{n}^{v}=\sum_{u\in S_{v}}w_{u,n}\cdot\mathbb{I}[u\in S_{v}^{n}],$$
where $S_{v}^{n}$ is the collection of vertices activated by information $I_{n}$ in $S_{v}$.
\end{enumerate}
This diffusion model partially enjoys submodularity. In a sense that the propagation range of each type of information as a function of seeds is submodular, which can be derived by applying Theorem 2 straightforwardly.

It is noticable that there has been a usable diffusion model for multiple information called \textbf{Competitive Linear Threshold Model} (CLTM). It functions in a similiar way as PLMMI except for that once the state of $S_{v}$ is fixed, the activation of $v$ is decided by comparing its threholds assigned to each $N$ information, with any tie broken in an arbitrary deterministic way. The proof and application in [12][13] readily adopted the way of proof as [1]. It is easy to see that CLTM and PLMMI introduce the identical distribution of activation over $\mathcal{V}$. 

One problem with competitive diffusion model is that: is there another choice other than CLTM or PLMMI? It is a little difficult to analyze this problem from the proof paradigm in \emph{live edges}. However, utilizing the Theorem 1 in our paper, we see that to ask for another diffusion model is tantamount to ask the $G_{v}$ for $v$ to be spanned on a basis other than those one-hot activation states (i.e., those $S'_{v}$ that corresponding to only one activation vertex in $S_{v}$ with one type of information, so the expectation of being activated by $I_{n}$ is a linear function of vertices activated by $I_{n}$ in $S_{v}$.) Unfortunately, \eqref{equation:11} indicates that \emph{only} the linear pattern is possible since for submodularity to be held for each type of information $I_{n}$, the inequality in \eqref{equation:1} has to degenerate to an equality so a linear combination is the only solution. 

Given the fact that we only welcome this linear pattern, we are now ready to turn an arbitrary competitive diffusion model into a submodular one. Recall that we can convert an arbitrary diffusion model $M$ (of a single type of information) into a submodular one by projecting $\textbf{a}_{M}(v)$ to a convex space spanned by $\textbf{M}$ and $\textbf{b}$. In competitive scenario, we only need to project $\textbf{a}_{M,n}(v)$ ($v$'s probability to be activated by $I_{n}$ under $S_{v}$'s activation states of $I_{n}$) to the convex space below: 
$$C_{n}=\left\{\textbf{M}\textbf{b}_{n}\right\},$$
where $\textbf{M}$ is the same as \eqref{equation:5}, but $\textbf{b}_{n}$ is now a vector of length $2^{|S_{v}|}$ with positive elements only in its first $|S_{v}|$ components, the rest components are zero. Moreover, to ensure normalization, we must have:
\begin{equation}
\label{equation:13}
\sum_{n=1}^{N}\textbf{1}^{\text{T}}\textbf{b}_{n}=1.
\end{equation}
To conclude, we project $\left\{\textbf{a}_{M,n}(v) \right\}_{n=1}^{N}$ to the product space of $C_{n}$s under \eqref{equation:13}, which is easily checked to be a convex space, so such a projection is always safe.

\section{Conclusion}
\label{section:5}
In this paper, we induce the equivalent condition for diffusion to be submodular. Compared with established paradigms that depend on specific models, our method is more general and concise. Moreover, we utilize the equivalent condition to coin a projection operator that maps an arbitrary diffusion model into a submodular one. At last, we propose a diffusion model for heterogeneous information that partially enjoys submodularity. Moreover, we adopt our theoretical way to show that this linear pattern in competitive diffusion is the only appropriate model that maintain submodularity. In the future, we are going to research competitive influence maximization, during which we can utilize the PLMMI we have just presented.

\section{References}
[1]. Kempe, David, Jon Kleinberg, and Éva Tardos. "Maximizing the spread of influence through a social network." Proceedings of the ninth ACM SIGKDD international conference on Knowledge discovery and data mining. ACM, 2003.

[2]. Nemhauser, George L., Laurence A. Wolsey, and Marshall L. Fisher. "An analysis of approximations for maximizing submodular set functions—I." Mathematical programming 14.1 (1978): 265-294.

[3]. Han, Meng, and Yingshu Li. "Influence analysis: A survey of the state-of-the-art." Mathematical Foundations of Computing 1.3 (2018): 201-253.

[4]. LESKOVEC J, KRAUSE A, GUESTRIN C, et al. Cost-effective outbreak detection in networks[C], Proceedings of the 13th ACM SIGKDD International Conference on Knowledge discovery and data mining. San Jose, CA, USA: ACM, 2007: 420–429.

[5]. Amit Goyal, Wei Lu, Laks V.S. Lakshmanan. CELF++: optimizing the greedy algorithm for influence maximization in social networks[C]. 2011.

[6]. Cormen T H, Leiserson C E, Rivest R L, et al. Introduction to Algorithms, Third Edition[M]. 2009.

[7]. Folland G B. Real analysis[M]. 1999.

[8]. Stephen Boyd, Lieven Vandenberghe. Convex Optimization[M], 2004.

[9]. Bharathi, Shishir, David Kempe, and Mahyar Salek. "Competitive influence maximization in social networks." International workshop on web and internet economics. Springer, Berlin, Heidelberg, 2007.

[10]. Carnes, Tim, et al. "Maximizing influence in a competitive social network: a follower's perspective." Proceedings of the ninth international conference on Electronic commerce. ACM, 2007.

[11]. Hui L, Bhowmick S S, Cui J, et al. GetReal: Towards Realistic Selection of Influence Maximization Strategies in Competitive Networks.[C] 2015.

[12]. A. Borodin, Y. Filmus and J. Oren. Threshold models for competitive influence in social networks. In WINE, pages 539-550, 2010.

[13]. Xinran He, Guojie Song, Wei Chen and Qingye Jiang. Influence Blocking Maximization in Social Networks under the Competitive Linear Threshold Model, SIAM, 463-474, 2011.

\end{spacing}
\end{document}